\documentclass[12pt]{article}

\begin{document}
	\thispagestyle{empty}
                \begin{flushright}
                hep-th/0007021\\
                \end{flushright}
               \vspace{1cm}
\begin{center}
Conformal symmetry of superstrings\\
on $AdS_3\times S^3\times T^4$ and
D1/D5 system\\

Abbas Ali\\

                {\it Physics Department }\\
                {\it Aligarh Muslim University }\\
                {\it Aligarh-202 002, India}\\
\end{center}

        \begin{abstract}
Conformal field theory of the D1/D5 system and superstrings on 
$AdS_3\times S^3\times T^4$ is studied with particular attention 
to the world-sheet fields corresponding to the $T^4$ part.  A solution to the
spacetime $N=4$ superconformal symmetry doubling and other problems
is proposed. It is argued that the relevant spacetime symmetry should 
be based on the middle $N=4$ superconformal algebra. 
It is discussed as to why this superconformal structure has been
missed so far.
      \end{abstract}

\vfil
e-mail: pht05aa@amu.up.nic.in

\newpage

\section{Introduction}

It was known for some time that there exist two commuting copies of infinite
dimensional conformal symmetry in any theory of three
dimensional gravity with negative cosmological constant
at the boundary of the spacetime\cite{brown-henneaux}.
Interest in this subject was renewed because of the AdS/CFT
correspondence enunciated in \cite{maldacena} and  made more precise later
\cite{gubser-klebanov-polyakov, witten}. This has lead to some remarkable 
advances, for example, in the study black holes\cite{btz}-\cite{strominger}
and of Wilson loops (see \cite{sonnenschein} for a review). Examination of
this correspondence from the world sheet, Green-Schwarz formalism and D1/D5
system points of view  has lead to even better understanding 
\cite{giveon-kutasov-seiberg}-\cite{david-mandal-wadia}.

A recurring theme in these developments is the presence of 
higher superconformal symmetries.  Particularly 
both small and large $N=4$ symmetries are known to occur.
This happens, for example, in case of superstrings moving 
on ${\cal M}= AdS_3\times S^3\times T^4$ and  
${\cal N} = AdS_3\times S^3\times S^3\times S^1$ manifolds,
respectively.

Though this study has clarified many issues but some important details have 
been missed. This manifests itself in the following curious problem. We 
know that a condition to get spacetime supersymmetry is the occurrence of $N=2$
world-sheet supersymmetry. This requires the existence of a conserved
$U(1)_R$ current on the world-sheet under which the supercurrents $G^\pm$
have charges $\pm 1$. It turns out that one can not tread this standard
route in case of the superstrings on ${\cal M}$. This was
pointed out in reference \cite{giveon-kutasov-seiberg}. It was observed there 
that this route leads to some serious problems of interpretation
of spacetime supersymmetry. It was found that this analysis 
gives two copies of Ramond sector supercharges. This leads, in turn, to
two small $N=4$ superconformal symmetries on the boundary of ${\cal M}$.
This is puzzling and unexpected. Normally only one such symmetry should exist.

One would like to understand this occurrence of two
copies of the spacetime boundary symmetry.
These two copies can not be regarded as left and right 
sectors of the spacetime theory due to several reasons.
The world-sheet and spacetime chiralities are related and therefore
a single chirality on the world-sheet can not lead to both chiralities on
the spacetime boundary. Moreover the other chirality on the
world-sheet will still give more spacetime symmetry. 
Other interpretations of the second copy of the small $N=4$
superconformal symmetry are equally inconvenient.

There is another puzzling feature of the standard world-sheet
free field analysis of the AdS/CFT correspondence in case of the manifold
${\cal M}$. Two $U(1)$ currents are available on the
spacetime boundary which can be used to generate the 
$SU(2)$ Kac-Moody symmetry required to generate the small $N=4$ superconformal
symmetry. There is an inbuilt asymmetry in the choice of
$U(1)$ currents because only one of them is chosen by internal structure.
There should be a natural explanation of this observation.

These problems are addressed in this note. It turns out that
all of these problems are related to each other.
A closer look at the structure of the background manifold and the 
field content of the theory leads to their resolution.
It is convenient to begin
with the free field analysis of the CFT of long string in the context of D1/D5
system or rather its S-dual NS1/NS5. Eventually one gets a single copy of the
{\it midlle} rather than two copies of the small $N=4$ superconformal algebra.
We argue that the standard analysis using world-sheet free field approach to 
strings propagating on the manifold ${\cal M}$ should lead to the same 
conclusion. We also discuss as to how this aspect has remained undetected.

Rest of this papers is organised as follows. In Section 2 we study the
CFT of the long string on the manifold ${\cal M}$ using the free field analysis
in the context of D1/D5 system. Comments on the usual world-sheet 
free field analysis are contained in Section 3. We end this paper with 
a discussion in Section 4.

\section{CFT of the long string}

In reference \cite{seiberg-witten}, apart from other things, the
D1/D5 system, constructed from branes in $R^6\times T^4$, was studied.
If the numbers of D1 and D5 branes are $Q_1$ and $Q_5$ respectively
then for large value of $Q_1$ the D1/D5 system is described
by a conformal field theory on the boundary of $AdS_3\times S^3$.
This is the conformal field theory of the large string.
The free field realization of the corresponding S-dual
theory, the NS1/NS5 system, was obtained in above reference.
We start with a review of these results and then take a closer look at the 
symmetry generated by the free fields corresponding to the $T^4$ part
of ${\cal M}$.

Expression for the generators of this conformal theory are 
given below.
\begin{eqnarray}
T&= &-\frac{1}{2} \partial S^\mu S^\mu -{j^aj^a\over Q_5}
-\frac{1}{2}\partial\phi\partial\phi +{Q_5-1 \over \sqrt{2Q_5}}
\partial^2\phi \nonumber\\  
G^\mu&= &{1\over\sqrt 2}\partial\phi S^\mu - {2 \over
\sqrt {Q_5}} \eta^a_{\mu\nu}j^aS^\nu +{1\over 6\sqrt {Q_5}}
\epsilon_{\mu\nu\rho\sigma}S^\nu S^\rho S^\sigma - {Q_5-1\over\sqrt
{Q_5}} \partial S^\mu \nonumber\\
J^a&= &j^a+\frac{1}{2}\eta^a_{\mu\nu}S^\mu S^\nu.\label{freefield1}
\end{eqnarray}
Here the indices $\mu, \nu, \cdots$ take four values from $0$ to $3$ and 
$a, b, \cdots$ take three values from $1$ to $3$. These operators satisfy the
following OPEs of the small $N=4$ superconformal algebra with $c=6(Q_5-1)$
\begin{eqnarray}
G^\mu(z)G^{\nu}(w) &\sim& {2\delta^{\mu\nu} (Q_5-1) \over
(z-w)^3} -{4 
\eta^a_{\mu\nu} J^a\over (z-w)^2} 
 + { \left(\delta^{\mu\nu}T - 2 \eta^a_{\mu\nu} \partial
J^a\right) \over z-w}\nonumber\\
J^a(z)J^b(w)& \sim& -{\delta^{ab} (Q_5-1)/2\over (z-w)^2} +
{\epsilon^{abc}J^c \over z-w}\nonumber\\
T(z)J^a(w)& \sim& {J^a \over (z-w)^2} + {\partial J^a \over z-w}~,
~~~~~~ J^a(z)G^\mu(w) \sim {\eta^a_{\mu\nu} \over z-w} G^\nu
\nonumber\\
T(z)T(w) &\sim& {3(Q_5-1) \over(z-w)^4} +{2T \over (z-w)^2} +
{\partial T\over z-w}
\nonumber\\
T(z)G^\mu(w) &\sim& {{3\over 2} G^\mu \over (z-w)^2} + {\partial G^\mu
\over (z-w)}
\label{small1}
\end{eqnarray}
where 't Hooft $\eta$ and ${\bar\eta}$ symbols are defined as 
\begin{eqnarray}
\eta^a_{\mu\nu}&= &
\alpha^{+a}_{\mu\nu}=\frac{1}{2} (\delta_{a\mu}\delta_{0\nu}-
\delta_{a\nu}\delta_{0\mu} +\epsilon_{a\mu\nu}),\nonumber\\
{\bar\eta}^a_{\mu\nu}&= &
\alpha^{-a}_{\mu\nu}=\frac{1}{2} (\delta_{a\nu}\delta_{0\mu} 
-\delta_{a\mu}\delta_{0\nu} +\epsilon_{a\mu\nu}).\label{eta}
\end{eqnarray}
OPEs (\ref{small1}) can be verified using the two point functions
\begin{eqnarray}
S^\mu(z)S^\nu(w) &\sim &-{\delta^{\mu\nu} \over z-w},
~~~~~~\partial\phi(z)\partial\phi(w) \sim  -{1 \over (z-w)^2}\nonumber \\
j^a(z)j^b(w) &\sim &-{\delta^{ab} (Q_5-2)/2\over (z-w)^2} +
{\epsilon^{abc}j^c \over z-w}.\label{2pt1}
\end{eqnarray}

Apart from this symmetry one may construct one more  set of
generators realizing another
small $N=4$ superconformal symmetry using the free fields
corresponding to the $T^4$ part of the manifold.
This realization will have a  central charge 
$c=6$. This takes the total of the central charge to $6Q_5$.

A convenient path to proceed further is to write down the generators of the
small $N=4$ corresponding to $T^4$ part of the manifold.
Since we require $c=6$ therefore putting $Q_5=2$
in (\ref{freefield1}) will give us one possible  realization.
Let us denote the corresponding ``free" fields by the symbols
$\varphi$, ${\tilde j}^a$ and ${\tilde S}^\mu$. Using these fields and 
$Q_5$ put equal to 2 one gets the following expressions for the
generators. 
\begin{eqnarray}
{\tilde T}&= &-\frac{1}{2} \partial{\tilde S}^\mu{\tilde S}^\mu 
-\frac{1}{2}{\tilde j}^a{\tilde j}^a 
-\frac{1}{2}\partial\varphi\partial\varphi
+\frac{1}{2}\partial^2\varphi
\nonumber\\  
{\tilde G}^\mu&= &{1 \over \sqrt 2}\partial\varphi{\tilde S}^\mu - 
{\sqrt 2}\eta^a_{\mu\nu}{\tilde j}^a{\tilde S}^\nu
-\frac{1}{6\sqrt2}\epsilon_{\mu\nu\rho\sigma}
{\tilde S}^\nu {\tilde S}^\rho {\tilde S}^\sigma
 - {1 \over \sqrt 2}\partial{\tilde S}^\mu\nonumber\\
{\tilde J^a}&= &{\tilde j}^a+
\frac{1}{2}\eta^a_{\mu\nu}{\tilde S}^\mu{\tilde S}^\nu
\label{freefield2}
\end{eqnarray}
The two point functions for the new fields are
\begin{eqnarray}
{\tilde S}^\mu(z){\tilde S}^\nu(w) &\sim &-{\delta^{\mu\nu} \over z-w},
~~~~~~\partial\varphi(z)\partial\varphi(w) \sim  -{1 \over (z-w)^2}\nonumber \\
{\tilde j}^a(z){\tilde j}^b(w) &\sim &
{\epsilon^{abc}j^c \over z-w}.\label{2pt2}
\end{eqnarray}

Though the operators (\ref{freefield2}) obey the small $N=4$ superconformal 
algebra with central charge ${\tilde c}=6$ but  
there are some serious shortcomings in this symmetry.
First of all the algebra so generated has an $SU(2)$
Kac-Moody symmetry. This is expected for the internal structure
of the $N=4$ symmetry but not  acceptable because $T^4$ does
not have an $SU(2)$ isometry. Even if we devise a way of avoiding
this problem then the $U(1)^4$ affine symmetry expected for the
present manifold $T^4$ is missing. Boson $\varphi$ is anomalous
and we do not have even a single independent $U(1)$ affine symmetry
apart from the one occurring as the Cartan subalgebra of $SU(2)$.
Another awkward feature is that the level of the $SU(2)$ currents 
${\tilde j}^a$ is zero.

First one of these problems, that is, the existence of an extra
$SU(2)$ is an indication that the algebra generated by the
generators (\ref{freefield2}) does not have an independent
existence. The fields occurring in eqn.(\ref{freefield2})
must be assimilated in the over all symmetry algebra. We shall 
do so in a shortwhile.

Two things have to be done to to solve the second problem. First of all we must
find a realization in which the boson $\varphi$ is not anomalous so that it can
give us a $U(1)$ current. Secondly we should drop the classical currents
${\tilde j}^a$ which are not serving very useful purpose here. In there place
we must introduce three more free bosons $\varphi^a$ to get three other $U(1)$
currents. Thus we will have four free bosons $\varphi^\mu$, such that 
$\varphi=\varphi^0$, and four free fermios ${\tilde S}^\mu$ which we must use
to get a free field realization of the small algebra with central charge 6.
Such a realization is easy to write is is already known. Expressions of the
generators in this realization are given below.
\begin{eqnarray}
{\check T}&= &-\frac{1}{2} \partial{\tilde S}^\mu{\tilde S}^\mu 
-\frac{1}{2}\partial\varphi^\mu\partial\varphi^\mu \nonumber\\  
{\check G}^\mu&= &{1 \over \sqrt 2}\partial\varphi{\tilde S}^\mu - 
{\sqrt 2}\eta^a_{\mu\nu}\partial\varphi^a{\tilde S}^\nu
\nonumber\\
{\check J^a}&= &
\frac{1}{2}\eta^a_{\mu\nu}{\tilde S}^\mu{\tilde S}^\nu.
\label{freefield3}
\end{eqnarray}
These generators realize a small $N=4$ superconformal symmetry
with central charge 6. In addition to that there are four $U(1)$ currents
given by $U^\mu\sim\partial\varphi^\mu$, all non-anomalous.
That is not all. A comparison with eqn.(16) of reference \cite{class} reveals
that eqs.(\ref{freefield3}) give a realization of the middle $N=4$
superconformal algebra, a symmetry more stringent than small
one\cite{middle1}-\cite{middle2}. This, of course, does not mean that we have 
a middle  superconformal 
symmetry on $T^4$. This is because this realization too has an $SU(2)$
Kac-Moody symmetry while there is no $SU(2)$ isometry on $T^4$.
(Of course this argument will apply only if the naive expectation that
the all the Kac-Moody symmetries on the AdS boundary should
come from an isometry. In this connection one should remember that AdS$_3$
part leads to the infinite Virasoro symmetry on the boundary of the
spacetime.)

The complete spacetime superconformal symmetry on ${\cal M}$
will be a suitable combination of the generators 
(\ref{freefield1}) and (\ref{freefield3}).
In this combination only the $SU(2)$ affine symmetry of (\ref{freefield1})
and $U(1)^4$ of (\ref{freefield3}) should be present not the
$SU(2)$ of (\ref{freefield3}). 
In other words our problem is reduced to the following one.
We have a set of field $\phi$, $\varphi^\mu$, $S^\mu$, ${\tilde S}^\mu$
and $SU(2)$ currents $j^a$. Using them we have to write a free field
realization which must have at least small $N=4$ superconformal symmetry,
an $SU(2)$ current algebra and four $U(1)$ symmetries. Answer to this problem
is already known. These conditions lead to a realization of the middle $N=4$
superconformal algebra. This is given in eqn.(23) of ref.\cite{class}.
After some rescalings and suitable redefinitions this realization
can be written as follows.
\begin{eqnarray}
\hat{T}(z)& = &-\frac{1}{2}(\partial\phi)^2
-\frac{Q_5-1}{\sqrt{2Q_5}}\partial^2\phi 
-\frac{1}{Q_5}j^aj^a-\frac{1}{2}\partial\varphi^\mu\partial\varphi^\mu
\nonumber\\
& &-\frac{1}{2}S^\mu\partial S^\mu 
-\frac{1}{2}{\tilde S}^\mu\partial{\tilde S}^\mu,\nonumber\\
\hat{U}^\mu(z)& = &\partial\varphi^\mu(z), ~~~~~\hat{Q}^\mu(z)=
\frac{1}{\sqrt 2}{\tilde S}^\mu(z),
\nonumber\\
\hat{J}^1(z)& = &j^1-\frac{1}{2}S^0S^1+\frac{1}{2}S^2S^3
-\frac{1}{2}{\tilde S}^0{\tilde S}^1
+\frac{1}{2}{\tilde S}^2{\tilde S}^3,{\rm cyclic ~for}~\hat{J}^{2,3},
\nonumber\\
\hat{G}^0(z)& = &\frac{1}{\sqrt 2}\partial\phi S^0-\frac{Q_5-1}{\sqrt{Q_5}}
\partial S^0
+\frac{1}{\sqrt 2}\partial\varphi^\mu{\tilde S}^\mu\nonumber\\
& &+\frac{1}{\sqrt{Q_5}}(j^aS^a+S^1S^2S^3),\nonumber\\
\hat{G}^1(z)& = &\frac{1}{\sqrt 2}\partial\phi S^1
-\frac{Q_5-1}{\sqrt{Q_5}}\partial S^1
+\frac{1}{\sqrt 2}\partial\varphi{\tilde S}^1 \nonumber\\
&{}&+\frac{1}{\sqrt{Q_5}}(-j^1S^0+j^2S^3-j^3S^2
-S^0S^2S^3)\nonumber\\
&{}&+\frac{1}{\sqrt 2}(-\partial\varphi^1{\tilde S}^0
+\partial\varphi^2{\tilde S}^3 -\partial\varphi^3{\tilde S}^2).
\label{middle4}
\end{eqnarray}
with cyclic expressions for $\hat{G}^2(z)$ and $\hat{G}^3(z)$.
These operators satisfy the OPEs of the middle $N=4$ superconformal algebra.
These OPEs include all the ones in eqn.(\ref{small1}) with $T$, $G$, $J^a$ and $Q_5-1$
replaced by $\hat{T}$, $\hat{G}$, $\hat{J}^a$ and $Q_5$ respectively.
Additional OPEs are given below.
\begin{eqnarray}
\hat{T}(z)\hat{U}^\mu(w)& = &\frac{\hat{U}^\mu}
{(z-w)^2}+\frac{\partial\hat{U}^\mu}{z-w},
\nonumber\\
\hat{T}(z)\hat{Q}^\mu(w)& = &\frac{\hat{Q}^\mu /2}
{(z-w)^2}+\frac{\partial\hat{Q}^\mu}{z-w},
\nonumber\\
\hat{U}^a(z)\hat{G}^\mu(w)& =& \frac{\eta^{a}_{\mu\nu}
\hat{Q}^\nu}{(z-w)^2},
~~~~~~\hat{U}^\mu(z)\hat{U}^\nu(w)= 
-\frac{\delta^{\mu\nu}}{(z-w)^2},\nonumber\\
\hat{J}^a(z)\hat{U}^\mu(w)& = &0=\hat{U}^\mu(z)\hat{Q}^\nu,
~~~~~~\hat{U}(z)\hat{G}^\mu(w)=\frac{\hat{Q}^\mu}{(z-w)^2},
\nonumber\\
\hat{Q}^\mu(z)\hat{G}^\nu(w)& = &
\frac{\eta^{a}_{\mu\nu}\hat{U}^a-\delta^{\mu\nu}\hat{U}/2}{z-w},
\nonumber\\
\hat{J}^a(z)\hat{Q}^\mu(w)& = &
\frac{\eta^{a}_{\mu\nu}\hat{Q}^\nu}{z-w}, 
~~~\hat{Q}^\mu(z)\hat{Q}^\nu(w)=-\frac{\delta^{\mu\nu}/2}
{z-w}.
\label{middle2}
\end{eqnarray}
Here $\hat{U}(z)=\hat{U}^0(z)$ and the central charge of the realization
is  $\hat{c}=6Q_5$.

Thus with this analysis we conclude that for the manifold ${\cal M}$
the D1/D5 system should be described by a middle rather than the small $N=4$
superconformal symmetry. Particularly we need not think of the symmetry to be 
consisting of two small $N=4$ parts corresponding to $AdS_3\times S^3$ and 
$T^4$ factors. 

\section{Comments on the world-sheet analysis of superstrings on ${\cal M}$}

 It is well known that superstring propagating on ${\cal N}$ lead to the 
large $N=4$ superconformal symmetry. In the usual world-sheet analysis this 
was shown in \cite{elitzur-feinerman-giveon-tsabar}. Another well known
fact is that In\"on\"u-Wigner contraction of the large $N=4$
superconformal symmetry leads to the middle one\cite{middle1,middle2}.
In the corresponding limit on the group manifold we
go from ${\cal N}$ to ${\cal M}$.
If we do a straightforward extrapolation of the results of the
analysis in the last section we again reach the same conclusion.
But the direct construction of the symmetry on the boundary of the
spacetime in case of the superstrings on ${\cal M}$
have so far lead to the small $N=4$ superconformal structure only.
How did we miss the middle $N=$ superconformal structure?
Then there is the serious problem of interpreting the results
of the standard world-sheet approach to spacetime supersymmetry described in 
the Appendix B of reference \cite{giveon-kutasov-seiberg}.

These two problems are complementary to each other.
If we assume that the actual spacetime superconformal
symmetry of the superstrings on ${\cal M}$ is the middle one
then the second problem has a natural solution.
In this section we take the first one and deal with the second one in 
the next section.

To answer the first problem raised above
we start with the following observation. The usual world-sheet
analysis of strings on $AdS_3$ manifolds gives us most easily the
global part of the spacetime superconformal symmetry.
Some of the (anti-) commutators of this algebra are
\begin{eqnarray}
\left[L_m, L_n\right]& = &(m-n)L_{m+n},~~
\left[L_m, \Phi_n\right] = [(d_\Phi-1)m-n]\Phi_{m+n},\nonumber\\
&{}&\Phi_n\in\{G^\mu_n, A^{\pm a}_n, U_n, Q^\mu_n\}, 
d_\Phi\in\{3/2, 1, 1, 1/2\},\nonumber\\
\{G^\mu_r, G^\nu_s\}& = & \delta^{\mu\nu}L_{r+s}
+2(r-s)[\gamma\alpha^{+a}_{\mu\nu}
A^{+a}_{r+s}+(1-\gamma)\alpha^{-a}_{\mu\nu}A^{-a}_{r+s}],
\nonumber\\
\left[A^{\pm a}_0, G^\mu_r\right]& = &
\alpha^{\pm a}_{\mu\nu}G^\nu_{r},~~
\left[A^{\pm a}_0, A^{\pm b}_0\right] =
\epsilon^{abc}A^{\pm c}_{0},
\left[A^{-a}_0, A^{+b}_0\right] = 0,
\nonumber\\
\left[U_0, G^\mu_r\right]& =& 0,~~
\left[U_0, A^{\pm a}_0\right] = 0, ~~~
\left[U_0, U_0\right] = 0
\label{large1}
\end{eqnarray}
and the rest of them are
\begin{eqnarray}
\{Q^\mu_r, G^\nu_s\}& = &\alpha^{+a}_{\mu\nu}A^{+a}_{r+s}
-\alpha^{-a}_{\mu\nu}A^{-a}_{r+s}+(\delta^{\mu\nu}/2)U_{r+s},\nonumber\\
\left[A^{\pm a}_0, Q^\mu_r\right]& = &\alpha^{\pm a}_{\mu\nu}Q^\nu_{r},
~~\left[U_0, Q^\mu_r\right]=0, 
\nonumber\\
\{Q^\mu_r, Q^\nu_s\}& =& -\frac{c}{12\gamma(1-\gamma)}
\delta^{\mu\nu}\delta_{r+s,0},
\label{large2}
\end{eqnarray}
where $A^{\pm a}$ are the two
$SU(2)$ currents and $\gamma$ is a parameter related to the levels of the
to $SU(2)$ algebras.

The standard analysis on the world-sheet takes us to the (anti-) commutators 
(\ref{large1})\cite{elitzur-feinerman-giveon-tsabar}. At the same time
this part misses the (anti-) commutators crucial for the
middle $N=4$ structure when one goes from the manifold
${\cal N}$ to ${\cal M}$. Particularly dimension half fermionic
generators are completely missing from eqn.(\ref{large1}).
Thus contraction of (\ref{large1}) will easily lead to
the small $N=4$ algebra but will not know about the remaining
structure relevant for the middle algebra. 
The small algebra is a subalgebra of the middle one and looking
at it will not lead to the larger algebra. Thus to get the
complete and real symmetry of the theory one has to explore
beyond the small algebra. One way to remedy the situation is to start
with the full $N=4$ superconformal algebra rather than the partial set
(\ref{large1}) of the (anti-) commutators. Then the middle $N=4$
superconformal algebra is obtained by an In\"on\"u-Wigner contraction
\cite{middle1,middle2}. At present this strategy can be implemented
for the Wakimoto type of realization\cite{ito}.

\section{Discussion}

A natural conclusion of the observations made in the earlier Sections
is that the 
spacetime symmetry for the superstrings moving on the manifold ${\cal M}$ and
the conformal field theory of the related D1/D5 should be based
on the middle $N=4$ superconformal algebra. This is relevant in its own right
because the $T^4$ part of the manifold should play its role in 
deciding the symmtery structure. In addition to that it should lead to a 
solution of the problems encountered in the
standard world-sheet analysis of superstrings on ${\cal M}$.

Doubling of the spacetime superconformal symmetry disappears if the generators
of one of the copies of the symmetry are rehashed to enlarge the other copy
to the level of middle algebra. Number counting is certainly in favour of
this proposal. The small algebra has eight generators while the middle one
has sixteen of them. Out of the eight spin fields in the Green-Schwarz
analysis four should lead to the four supercharges and four to the
other four fermionic dimension half generators. (Same thing
should happen even in the case of the superstrings moving on ${\cal N}$
regarding the expressions for the dimension half generators.)
The problem of asymmetry in selection of the $U(1)$ current, mentioned in the 
Introduction, also has a natural explanation. The $U(1)$ not required
for the extension to $SU(2)$ level is one of the four $U(1)$ currents
required for the middle algebra. Question of chirality mixing does not arise
at all.

An open problem in this regard is the explicit construction of the
generators relevant for the middle structure in the same way as it has been 
done for the other generators in the old fashioned world-sheet analysis.
Problem for the D1/D5 system was easy because things really came down
to the level of simple free field theories if we use the approach
pioneered in reference \cite{seiberg-witten}.
For the superstrings on ${\cal M}$ or ${\cal N}$ the problem is compounded
by the difficulty in  the construction of the dimension half generators
using the spin fields. It will be more illuminating if one can find a
technique to promote the spin fields to the level of free fields on the
AdS$_3$ boundary parallel to the approach in case of D1/D5 system.
It certainly will be more transparent than the standard procedure of 
bosonosing the spin fields and then constructing fermionic operators out 
of the bosons thus obtained.

Next step in this work will be to analyze the spectrum of the superstrings 
moving on ${\cal M}$ in the light of the symmetry described above. Even before 
that one should study the representations of the middle $N=4$ superconformal 
symmetry itself. We hope to return to these issues in future.

{\it Acknowledgements.} I am thankful to Profs. S.K.~Singh and Hashim Rizvi
for encouragement, Dr. Sudhakar Panda for reading the manuscript and Prof.
H.S.~Mani for hospitality at the Mehta Research Institute, Allahabad where
part of this work was completed.

\end{document}